\documentstyle[amsfonts,preprint,aps]{revtex}
\begin{document}
\title{Realization of the Optimal Universal Quantum Entangler}
\author{Fabio Sciarrino$^{\ast }$, Francesco De Martini$^{\ast }$ and Vladimir Bu%
\v{z}ek$^{\dagger }$}
\address{$^{\ast }$Dipartimento di Fisica and Istituto Nazionale per la Fisica della\\
Materia,\\
Universita' di Roma ''La Sapienza'', Roma, 00185 Italy,\\
$\dagger $ Research Center for Quantum Information, Institute of Physics,\\
Slovak Academy of Sciences\\
D\'ubravsk\'a cesta 9, 845 11 Bratislava, Slovakia.}
\date{\today}
\maketitle

\begin{abstract}
We present the first experimental demonstration of the ''optimal'' and
''universal'' quantum entangling process involving qubits encoded in the
polarization $(\overrightarrow{\pi })$ of single photons. The structure of
the ''quantum entangling machine'' consists of the {\it quantum injected}
optical parametric amplifier by which the contextual realization of the $%
1\rightarrow 2$ universal quantum cloning and of the universal NOT (U-NOT)\
gate has also been achieved.
\end{abstract}


The two distinctive features of quantum kinematics are the {\em quantum
superposition principle} and the {\it quantum entanglement}. Among many
consequences of the statistical character of quantum kinematics, a {\em %
complete} determination of the unknown state of a quantum system can\ be
attained only when a complete measurement (i.e. the measurement of the
quorum of observables) is performed on an {\em infinite} ensemble of
identically prepared quantum objects. The measurement on a finite ensemble
results in an imperfect reconstruction of the quantum state \cite{1}.

If we consider the physical world to be represented by states of quantum
objects then it is obvious that the quantum information (QI)\ processing is
fundamentally different from any processing on a classical level. One of the
main differences is that, in general, given just one physical object
carrying a specific quantum information this one cannot be determined. In
addition many operations on individual quantum objects prepared in unknown
quantum states cannot be performed perfectly. A renowned example of such a
constraint is the impossibility of cloning (copying) an unknown quantum
state $|\Psi \rangle $ \cite{2}, i.e. a {\it universal} machine realizing
exactly the transformation $|\Psi \rangle |0\rangle \rightarrow |\Psi
\rangle |\Psi \rangle ,$ being $|0\rangle $ a known state of the copier,
cannot exist. On the other hand an approximate, i.e. {\it optimal},
universal quantum cloning machine has been theoretically proposed~\cite{3}
and experimentally realized \cite{4,5}. Another relevant example of an
impossible task is the ''flipping'' of unknown qubits~\cite{6,7}, i.e. the\
realization of a {\it universal NOT-gate} operation $|\Psi \rangle
\rightarrow |\Psi ^{\perp }\rangle $ being: $\langle \Psi |\Psi ^{\perp
}\rangle =0$. The impossibility of flipping an unknown qubit has several
interesting consequences. For instance, it has been shown that encoding
information about unknown spatial spin orientation into parallel and
antiparallel pairs of qubits is different. Specifically, more information is
contained in the antiparallel spins. This purely quantum mechanical (QM)\
effect is due to the entanglement that appears in the process of optimal
measurement. However, in spite of this constraint an optimal universal NOT
gate has indeed been proposed \cite{6} and realized \cite{8}.

To pursue at a deeper level this most significant quantum-classical endeavor
consider here the central role of state-entanglement in quantum mechanics.
As it is well known this fundamental physical condition, pervasive of the
entire QI\ domain, is the key ingredient of all quantum nonlocality tests
involving either Bell inequalities or Hardy's ''ladder proofs''\cite{9,10}.
Furthermore, it lies at the core of important QI protocols as quantum
teleportation, dense coding, etc. Following the above reasonings, one may
ask then again whether it is possible to realize exactly the map $|\Psi
\rangle |\Phi \rangle \rightarrow (|\Psi \rangle |\Phi \rangle +|\Phi
\rangle |\Psi \rangle )$ which implies the entanglement of two quantum
systems initially prepared in two unknown states $|\Psi \rangle $ and $|\Phi
\rangle $. Alternatively, the same question can be raised for the relevant
map $|\Psi \rangle \rightarrow (|\Psi \rangle |\Psi ^{\perp }\rangle +|\Psi
^{\perp }\rangle |\Psi \rangle )/\sqrt{2}\equiv |\{\Psi ,\Psi ^{\perp
}\}\rangle $ implying the ``translation'' of the information originally
encoded in any unknown state $|\Psi \rangle $ into the corresponding
entangled Bell state. This question has been addressed in Ref.\cite{11}
where it is shown that again the {\it perfect} entangling transformation is
generally impossible but, once again an approximate universal entangling
machine can be designed. In addition and most interestingly, it can be shown
that any {\it optimal universal} quantum entangler, i.e. the one maximizing
the average fidelity of success, is realized within a combined, simultaneous
realization of the optimal universal quantum cloning and of the optimal
universal spin-flipping processes. In the present work we report the first
experimental demonstration of the ''optimal'' and ''universal'' quantum
entangling process within such a complex conceptual and experimental
framework.

Let us assume that QI is encoded in the polarization $(\overrightarrow{\pi })
$ of single photons. The structure of the ''quantum entangling machine''
and, contextually, of\ the $N=1$ to $M=2$ universal quantum cloning machine
and of the universal NOT (U-NOT)\ gate is the quantum injected optical
parametric amplifier (QI-OPA) \cite{5,12}. The action of this rather complex
machine can be described by the covariant transformation \cite{3}: 
\begin{equation}
|\Psi \rangle |\downarrow \rangle _{C}|\downarrow \rangle
_{AC}\Longrightarrow \sqrt{2/3}|\Psi \rangle |\Psi \rangle |\Psi ^{\perp
}\rangle _{AC}-\sqrt{1/3}|\{\Psi ,\Psi ^{\perp }\}\rangle |\Psi \rangle _{AC}
\end{equation}
where the first (unknown) state vector $|\Psi \rangle $ in the left-hand
side of \ the equation corresponds to the input, the second state vector
describes the system on which the information will be copied (''blank''
qubit), represented by the ''{\it cloning channel}'' (C), i.e. the injection
mode $k_{1}$, while the third state vector, the ''{\it anticloning channel}%
'' (AC), represents the state of the machine. Precisely, the state of the
machine is a qubit associated with the AC mode $k_{2}$. The blank qubit and
the cloner are initially in the known ground state $|\downarrow \rangle $.
At the output of the machine we find the completely symmetrized state $%
|\{\Psi ,\Psi ^{\perp }\}\rangle $ and two cloned qubits in the C channel: $%
\rho =$ $2/3|\Psi \Psi \rangle \langle \Psi \Psi |+$ $1/3|\{\Psi ,\Psi
^{\perp }\}\rangle \langle \{\Psi ,\Psi ^{\perp }\}|$. The density operator $%
\rho $ describes the best possible approximation of the perfect entangled
state $|\{\Psi ,\Psi ^{\perp }\}\rangle $. The most attractive feature of
this entangling machine is that the fidelity of its performance, i.e. the
distance between the output and the ideally entangled-state, does not depend
on the input state $|\Psi \rangle $ and takes the constant value $F=1/3.$
The machine itself after the cloning transformation is in the state $\rho
_{AC}=1/3|\Psi ^{\perp }\rangle \langle \Psi ^{\perp }|+$ $1/3\times {\bf I}$%
{\bf \ }, where ${\bf I}$ is the unity operator. This last density operator
is the best possible approximation of the spin-flip (U-NOT) operation
permitted by the quantum mechanics.

The symmetrization process was experimentally realized in a $2\times 2$
dimensional Hilbert space of photon polarization $(\overrightarrow{\pi })$%
{\bf \ }simultaneously with the realization of the linearized $N=1$, $M=2$
cloning process. \ Consider first the case of an input $\overrightarrow{\pi }
$-encoded qubit $\left| \Psi \right\rangle _{in}\ $associated with a single
photon with wavelength (wl) $\lambda $, injected on the input mode $k_{1}$
of the QI-OPA, the other input mode $k_{2}$ being in the vacuum state \cite
{12}. As for previous works, the photon was injected into the a nonlinear
(NL) BBO ($\beta $-barium-borate) 1.5 mm thick crystal slab, cut for Type II
phase matching and excited by a sequence of UV\ mode-locked laser pulses
having duration $\tau \approx $140 $f\sec $ and wl $\lambda _{p}$. The
relevant modes of the NL 3-wave interaction driven by the UV pulses
associated with mode $k_{p}$ were the two spatial modes with wave-vector
(wv) $k_{i}$, $i=1,2$, each supporting the two horizontal $(H)$ and vertical 
$(V)$ {\it linear}-$\overrightarrow{\pi }$'s of the interacting photons. The
QIOPA was $\lambda $-degenerate, i.e. the interacting stimulated emitted
photons had the same wl's $\lambda =%
%TCIMACRO{\UNICODE[m]{0xbd}}%
%BeginExpansion
{\frac12}%
%EndExpansion
\lambda _{p}=795nm$. The NL\ crystal orientation was set as to realize the
insensitivity of the amplification quantum efficiency $(QE)\;$to any input
state $\left| \Psi \right\rangle _{in}$, i.e. the {\it universality} (U)\ of
the entangling machine. It\ is well\ known that this key property is assured
by the squeezing hamiltonian \cite{12}:\ $\widehat{H}_{int}=i\chi \hbar
\left( \widehat{a}_{\Psi }^{\dagger }\widehat{b}_{\Psi \perp }^{\dagger }-%
\widehat{a}_{\Psi \perp }^{\dagger }\widehat{b}_{\Psi }^{\dagger }\right)
+h.c.$. The field operators sets $\left\{ \widehat{a}_{\Psi }^{\dagger },%
\widehat{a}_{\Psi }\right\} $, $\left\{ \widehat{a}_{\Psi \perp }^{\dagger },%
\widehat{a}_{\Psi \perp }\right\} $, $\left\{ \widehat{b}_{\Psi }^{\dagger },%
\widehat{b}_{\Psi }\right\} $ and $\left\{ \widehat{b}_{\Psi \perp
}^{\dagger },\widehat{b}_{\Psi \perp }\right\} $ refer to two mutually
orthogonal $\overrightarrow{\pi }$-states, $\left| \Psi \right\rangle $ and $%
\left| \Psi ^{\perp }\right\rangle $, realized on the two interacting
spatial modes $k_{1}$ and $k_{2}$ acted upon by the $\widehat{a}$ and $%
\widehat{b}$ operators, respectively. The $SU(2)$ invariance of $\widehat{H}%
_{int}$ implied by the U-condition, i.e. the independence of the OPA
``gain'' $g\equiv \chi t$ to any unknown $\overrightarrow{\pi }${\bf - }%
state of the injected qubit, $t$ being the interaction time, allows the use
of the subscripts $\Psi $ and $\Psi ^{\perp }$ in Eq.(1) \cite{12}.

The QIOPA apparatus adopted in the present work was arranged in the
self-injected configuration\ shown in Figure 1. The UV pump beam,
back-reflected by a spherical mirror $M_{p}$ with 100\% reflectivity and $%
\mu -$adjustable position ${\bf Z}$, excited the NL crystal in both
directions $-k_{p}$ and $k_{p}$, i.e. correspondingly oriented towards the
right hand side and the l.h.s. of Fig.1. A Spontaneous Parametric Down
Conversion (SPDC) process excited by the $-k_{p}$ UV\ mode created {\it %
singlet-states} of photon polarization $(\overrightarrow{\pi })$. The photon
of each SPDC pair emitted over $-k_{1}$ was back-reflected by a spherical
mirror $M$ into the NL crystal and provided the $N=1$ {\it quantum injection}
into the OPA excited by the UV beam associated with the back-reflected mode $%
k_{p}$. Because of the low pump intensity, the probability of the unwanted $%
N=2$ injection has been estimated to be $10^{-2}$ smaller than the one for $%
N=1$. The twin SPDC\ photon emitted over mode $-k_{2}$ , selected by the
devices (Wave-Plate + Polarizing Beam Splitter: $WP_{T}\ $+ $PBS_{T}$) and
detected by $D_{T}$, provided the ''trigger'' of the overall conditional
experiment. The three fixed quartz plates $(Q)$ inserted on the modes $k_{1}$%
, $k_{2}$ and $-k_{2}$ provided the compensation for the unwanted walk-off
effects due to the birefringence of the NL crystal. An additional walk-off
compensation into the BBO\ crystal was provided by the $\lambda /4$ WP
exchanging on mode $-k_{1}\ $the $\left| H\right\rangle $ and $\left|
V\right\rangle \ \overrightarrow{\pi }-$ components of the injected photon.
Because of the EPR non-locality of the emitted singlet, the $\overrightarrow{%
\pi }$-selection made on $-k_{2}$ implied deterministically the selection of
the input state $\left| \Psi \right\rangle _{in}$ on the injection mode $%
k_{1}$. All adopted photodetectors $(D)$ were equal SPCM-AQR14 Si-avalanche
single photon units with $QE^{\prime }s\cong 0.55$. One interference filter
with bandwidth $\Delta \lambda =6nm$ was placed in front of each $D$.

Since the U-condition of the apparatus was already tested in previous
experiments \cite{4,5} we limited ourselves to inject only one polarization
state on the input mode $k_{1}$, i.e. $\left| \Psi \right\rangle
_{in}=\left| H\right\rangle =\left| 1,0\right\rangle _{k_{1}}\otimes \left|
0,0\right\rangle _{k_{2}}$ where $\widehat{a}_{\Psi }^{\dagger }\left|
0,0\right\rangle _{k_{1}}=\left| 1,0\right\rangle _{k_{1}}$ and $\left|
m,n\right\rangle _{k_{1}}$ represents a product state with $m$ photons of
the mode $k_{1}$ having the polarization $\Psi =H$, and $n$ photons having
the polarization $\Psi ^{\perp }=V$. \ Assume the input mode $k_{2}$ to be
in the {\it vacuum state}. The initial $\overrightarrow{\pi }$-state\
evolves according the unitary operator $\widehat{{\bf U}}\equiv \exp \left(
-i\widehat{H}_{int}t\right) $: 
\begin{equation}
\widehat{{\bf U}}\left| \Psi \right\rangle _{in}\simeq \left|
1,0\right\rangle _{k1}\otimes \left| 0,0\right\rangle _{k2}+g\left( \sqrt{2}%
\left| 2,0\right\rangle _{k1}\otimes \left| 0,1\right\rangle _{k2}-\left|
1,1\right\rangle _{k1}\otimes \left| 1,0\right\rangle _{k2}\right) 
\label{OutputOPA}
\end{equation}
The above linearization procedure representing the 1st-order approximation
for the${\Bbb \ }$pure output state vector $\left| \Psi \right\rangle
_{out}\ $for $t>0,$ i.e. the restriction to the simplest $1\rightarrow 2$
cloning case, is justified here by the small experimental value of the {\it %
gain}: $g\approx 0.1$ \cite{5}. The first term in the expression $\propto g$
in Eq.(2) expresses the simultaneous emission on mode $k_{1}$ of the $%
1\rightarrow 2\;$cloned state $\left| 2,0\right\rangle _{k_{1}}$
corresponding to the state $\left| \Psi \Psi \right\rangle $ expressed by
the general theory and on mode $k_{2}$ of the {\it flipped} version of the
input qubit realizing the quantum U-NOT gate \cite{5}. The second term
expresses the emission on mode $k_{1}$ of the symmetrized state $\left|
1,1\right\rangle _{k_{1}}$under the present investigation. The two photons
emitted over the mode $k_{1}$ impinged on a balanced beamsplitter $(BS_{1})$
that coupled the mode $k_{1}$ to the output modes $a$ and $b.$ We restrict
our analysis to the cases in which the two photons emerge from different
output ports of the beamsplitter. The first term in $g$ in Eq.(2) hence
leads to the following normalized output state 
\begin{equation}
\left| \Psi \right\rangle _{out}=\sqrt{\frac{2}{3}}\left| H\right\rangle
_{a}\left| H\right\rangle _{b}\left| V\right\rangle _{k2}-\frac{1}{\sqrt{6}}%
\left( \left| H\right\rangle _{a}\left| V\right\rangle _{b}+\left|
V\right\rangle _{a}\left| H\right\rangle _{b}\right) \left| H\right\rangle
_{k2}
\end{equation}
This state was analyzed by the simultaneous excitation of the two detector
pairs $(a$ and $b)$ coupled respectively by the two $\overrightarrow{\pi }-$%
analyzers $PBS_{a}$ and $PBS_{b}$ to the two output modes of the Beam
Splitter $(BS_{1})$, and of the detector pair $(2)$ associated to the
polarizing beam splitter $PBS_{2}$ (Fig. 1). The histogram shown in Fig. 2
reports the experimental realization of the output state $\propto g$ by
expressing the probabilities of the various simultaneous state contributions
in Eq.(3). The variable $XYZ$\ of the histogram reads as follows: $X=$
polarization $\overrightarrow{\pi }-state$ detected by the detector pair $(a)
$ on the mode $k_{1}$, $Y=\overrightarrow{\pi }-state$ detected by the
detector pair $(b)$ on $k_{1}$, $Z=\overrightarrow{\pi }-state$ detected on
the mode $k_{2}$. The experimental values are\ found in good agreement with
the theoretical ones. The state probabilities related to the histogram
variables $VHH$ and $HVH$, i.e. detected in coincidence with a $\left|
H\right\rangle $ state realized on mode $k_{2}$, correspond precisely to the
realization of the two interfering terms of the bipartite entangled state $%
\left| \Phi \right\rangle _{out}=2^{-1/2}\left( \left| H\right\rangle
_{a}\left| V\right\rangle _{b}+\left| V\right\rangle _{a}\left|
H\right\rangle _{b}\right) $ over the modes $a$ and $b$. However the
existence of the contributions $VHH$ and $HVH$ alone is not a sufficient
proof of the entanglement feature, since the above observation is also in
agreement with a statistical mixture of $VHH$ and $HVH$. To demonstrate the
coherent superposition of the two terms we further performed a polarization
measurement in the $45^{\circ }$ basis on the modes $a$ and $b$ by rotating
the half-wave plates $WP_{a}$ and $WP_{b}$ by $22.5^{\circ }$. In this basis 
$\left| \Phi \right\rangle _{out}$ is expressed as $2^{-1/2}\left( \left|
+45^{\circ }\right\rangle _{a}\left| +45^{\circ }\right\rangle _{b}+\left|
-45^{\circ }\right\rangle _{a}\left| -45^{\circ }\right\rangle _{b}\right) $
where $\left| \pm 45^{\circ }\right\rangle =2^{-1/2}\left( \left|
H\right\rangle \pm \left| V\right\rangle \right) .$ We measured the
polarization correlation between the photons $a$ and $b$ with a
four-coincidences scheme involving the detectors $\left(
D_{T},D_{2},D_{a},D_{b}\right) $. The correlation measurement was compared
with the configuration in which there was no temporal overlap between the
injected photon and the back reflected UV pump. In this case the two
detected photons over the modes $a$ and $b$ have no correlation in the $%
45^{\circ }$ basis and hence there is the same probability for the photons
to have the same or different polarization. By moving the mirror $M_{p}$ in
order to continuously reach the temporal superposition, we should observe an
increase of the coincidence counts by a factor $R=2$ for the position $Z=0$
(Fig. 3). Fitting the experimental data with a gaussian function, we
estimate $R=1.68\pm 0.07$. We note that the peak of Fig.3 does not arise as
an amplification process since the component $\left| H\right\rangle _{k2}$
is not amplified (see Ref.~\cite{12,4}), instead it must be interpreted as a
consequence of the mode coalescence of two photons with orthogonal
polarization.

In conclusion we have experimentally demonstrated that the universal NOT
gate process lies at the basis of any universal entangling device. The
experiment enlightens the significance of the transformation Eq.(2) that
contextually implements, in a unifying manner, the universal NOT gate, the
universal optimal quantum cloning and the universal quantum entangler.
Indeed the optimality and the universality of the entangling process is
found to arise as a consequence of the same properties characterizing the
cloning and the spin-flipping processes \cite{13}. Finally note that the
optimal quantum entangler here realized in a more general NL context by a\
Optical Parametric Amplifier can also be implemented by a linear
state-symmetrization procedure involving the simultaneous realization of the
optimal quantum processor by a general modified quantum teleportation scheme 
\cite{14,11}.

We thank Daniele Pelliccia for early collaboration in the experiment. This
work has been supported by the FET European Network on Quantum Information
and Communication (Contract IST-2000-29681: ATESIT), the Marie Curie
Research and Training Network CONQUEST (Contract MRTN-CT-2003-505089) and by
PRA-INFM\ 2002 (CLON).

\centerline{\bf Figure Captions}

\vskip 8mm

\parindent=0pt

\parskip=3mm

Figure.1. Schematic diagram of the {\it self-injected} Optimal Parametric
Amplifier. The {\it universal optimal quantum entangler }is\ realized on the
cloning (C)\ channel (mode $k_{1}$). Micrometric adjustments of the
coordinate {\bf Z} of the UV\ mirror $M_{p}$ ensured the time superposition
in the active NL\ crystal of the UV\ 140 femtosecond pump pulses and of the
single photon pulse injected via back reflection by the fixed\ mirror $M$.

Figure.2. Probability distribution for the variables $XYZ$\ where $X,Y$ and $%
Z$ are the polarization $\overrightarrow{\pi }-state$ detected,
respectively, by the detector pair $(a)$ on the mode $k_{1}$, the detector
pair $(b)$ on the mode $k_{1}$, and the detector pair $(2)$ on the mode $%
k_{2}$. Each correlation data has been measured in a time of $2400s$.

Figure.3. Coincidence counts $\left( D_{T},D_{2},D_{a},D_{b}\right) $ versus
the position $Z$ of the UV mirror $M_{p}.$ The enhancement in the
coincidence counts is a signature of the entanglement of the state $\left|
\Phi \right\rangle _{out}$.

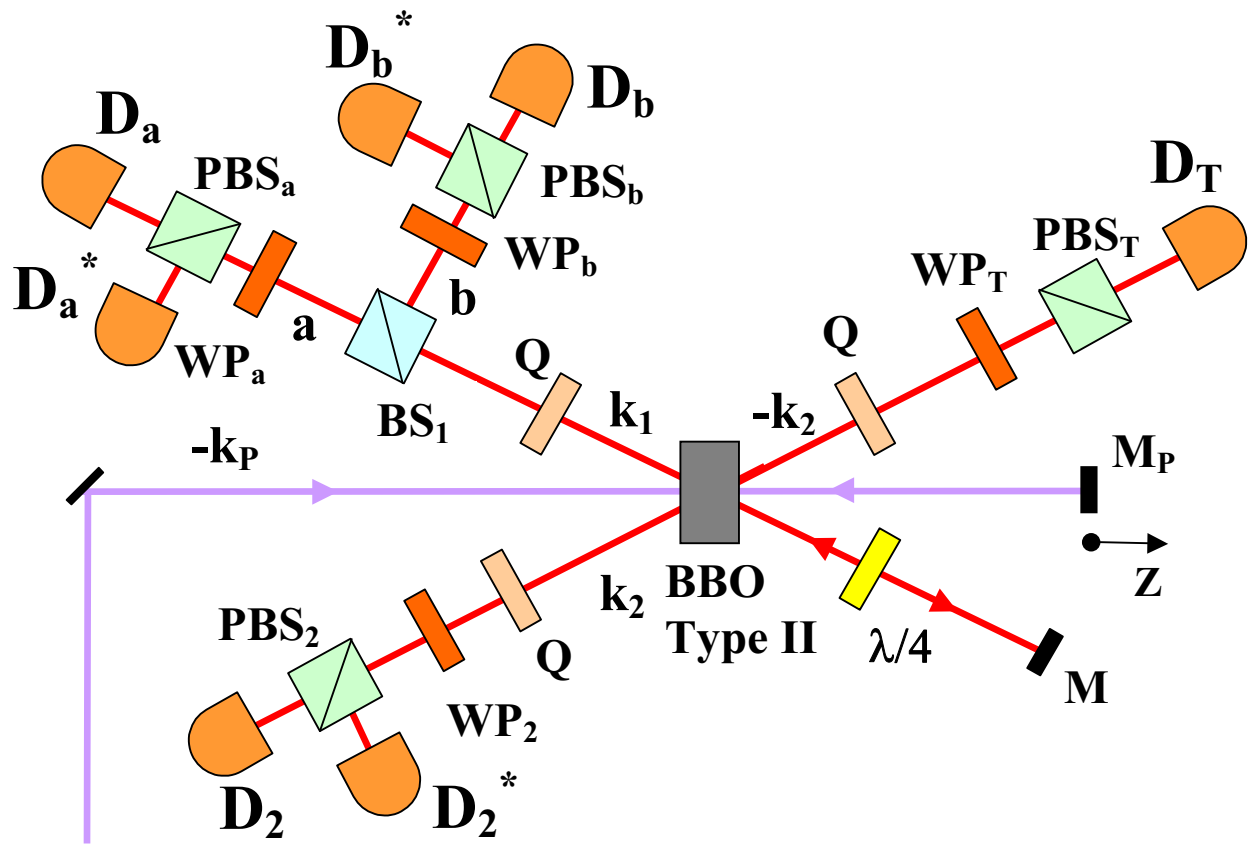
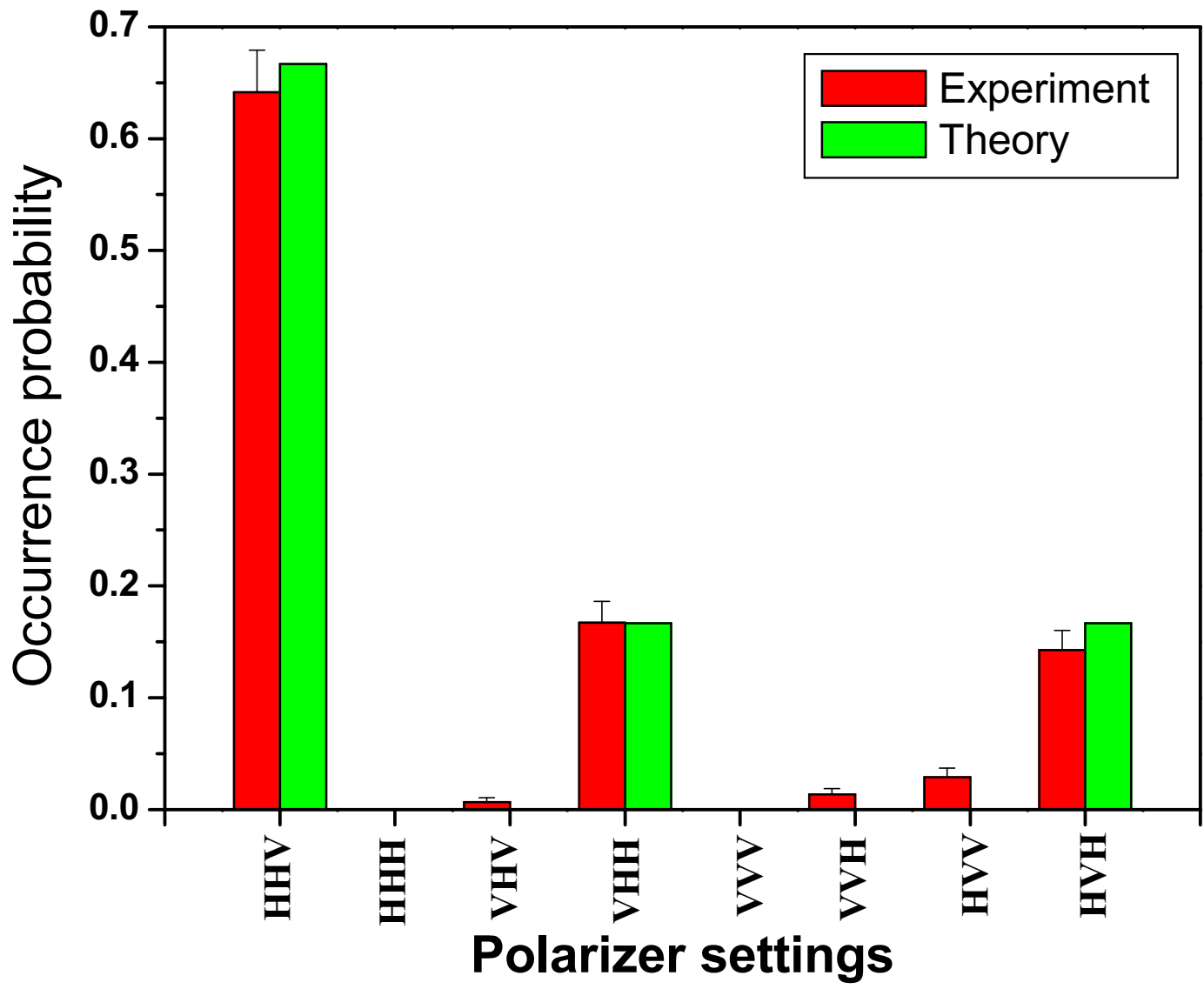
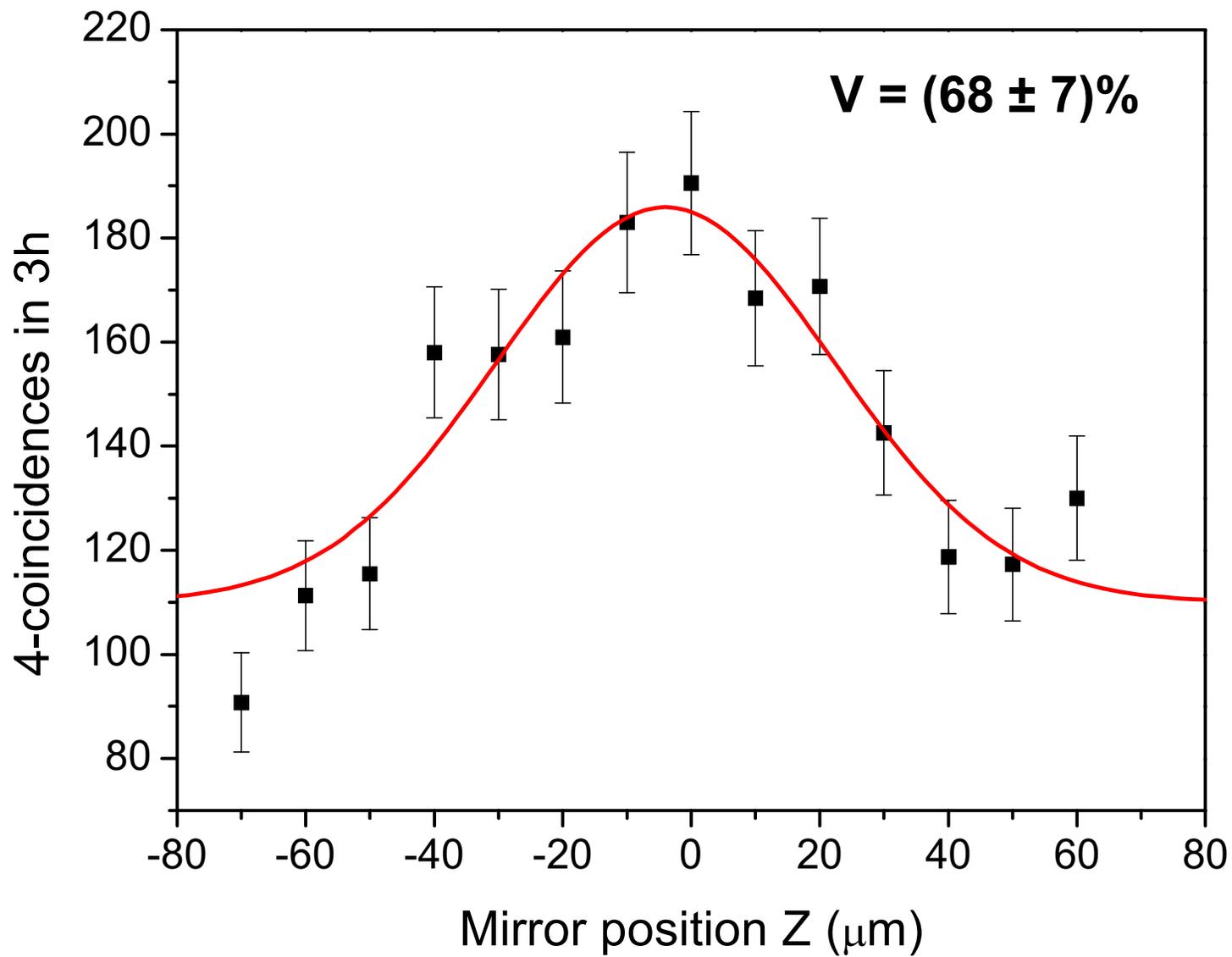
\end{document}